\documentclass[12pt]{article}

\begin{document} 
\bibliographystyle{unsrt} 
 
\begin{center} 
{\LARGE \bf Constraint Superalgebras and Their Application to Gauge Field Theories and String Theories}\\[10mm] 
 
Sultan Catto$^{\dag}$\\{\it Physics Department\\ The Graduate School and University Center\\ The City University of New York \\365 Fifth Avenue\\
New York, NY 10016-4309\\ and \\ Center for Theoretical Physics \\The Rockefeller University \\
1230 York Avenue\\ New York NY 10021-6399}\\[6mm] 
\end{center} 
\vbox{\vspace{5mm}}

\begin{abstract} We show that with every classical system possesing first class constraints that form a natural Lie algebra, we can asssociate a superalgebra that admits the constraint Lie algebra as a subalgebra. An odd generator of this superalgebra that commutes with the constraints is shown to be the BRST operator whose form follows from a non linear coset representation of the superalgebra. We further show the existence of the superalgebra for all Yang-Mills theories and for 26-dimensional bosonic strings.
\end{abstract}
\vbox{\vspace{5mm}}

PACS numbers: 12.40.Aa, 12.40.Qq, 11.30.Pb

\vbox{\vspace{10mm}}

$^\dag$ Work supported in part by DOE contracts No. DE-AC-0276 ER 03074 and 03075, and PSC-CUNY Research Awards.
\newpage

\section{Introduction}
Superalgebras were discovered in the context of hadronic physics, first as a generalization of the $SU(6)$ symmetry mixing states of different spin$^{\cite{miy}}$ and then by Ramond$^{\cite{ram}}$, Neveu and Schwarz$^{\cite{nev}}$ in the
construction of dual resonance models that accounted for parallel mesonic and baryonic Regge trajectories. Finally a class of supersymmetric relativistic field theories were discovered by Wess and Zumino$^{\cite{wess}}$ following initial examples by Gol'fand and Likthman$^{\cite{golfand}}$, and Akulov and Volkov$^{\cite{akulov}}$.  Hadronic
supersymmetry had two offshoots.  One was a fundamental superstring theory of particles$^{\cite{yon}}$, the other was the
approximate supersymmetry between even and odd nuclei developed as a model for nuclear structure, generalizing the $SU(6)$ interacting boson model$^{\cite{iac}}$, by Balantekin, Bars and Iachello$^{\cite{bal}}$.  

In this paper we shall deal with a totally different application of superalgebras to the quantization of Hamiltonian systems with first class constraints, following along the lines first introduced by Bowick and G\"ursey$^{\cite{bow}}$. We shall show that with every classical system possessing first class constraints that form a natural Lie algebra, we can associate a superalgebra that admits the constraint Lie algebra as a subalgebra.  An odd generator of this superalgebra that is invariant under the subalgebra of first class constraints is the BRST operator that was introduced in physics by Becchi, Rouet and Stora and also independently by Tyutin in the context of the covariant quantization of gauge theories$^{\cite{bec}}$.  The remaining odd operators that transform like the adjoint representation of the Lie superalgebra are the Faddeev-Popov ghost operators$^{\cite{fad}}$.  The BRST operator Q can also be expressed in terms of the first class constraints $\Phi_{i}$, the ghost operators $c^{i}$ and their canonical conjugate antighost operators $b_{i}$.  The resulting expression is a cubic function of those operators.  It was first written explicitly by Koszul$^{\cite{kos}}$ as the form of the cohomology operator of Lie groups as a step in Cartan's program for a complete differential geometric interpretation of Lie groups.  This was the same year in which Dirac$^{\cite{dir}}$ had formulated his theory of constrained Hamiltonian systems and their quantization.  Later on physicists rediscovered Koszul's formula.  The connection between the cohomology approach and the quantization of constrained systems approach was realized gradually during the seventies$^{\cite{fra}}$.  Looking back we see the simultaneous discovery of the two theories as yet another example of a happy conjuncture of great physical and mathematical ideas.

In physics Dirac's method was applied with the greatest success to the covariant quantization of non-abelian gauge
theories$^{\cite{sla}}$ and string theories$^{\cite{kat}}$.  The algebraic structure of the constraint-ghost system
was analyzed by many authors, including Fradkin, Vilkovisky and their collaborators$^{\cite{bat}}$, Marnelius$^{\cite{mar}}$, Schwarz$^{\cite{sch}}$, Baulieau$^{\cite{bau}}$, and Aldaya and collaborators$^{\cite{ald}}$. Recently there has been a flood of papers dealing with the BRST formalism in string theory and string field theories
that will not be the object of our main concern. We shall concentrate on the imbedding of the constraint Lie algebra in a superalgebra and give some applications. For some applications of the superalgebra method introduced in reference $\cite{bow}$ to string field theory we refer to articles in cited in reference ${\cite{cas}}$ among many others.

\section{The BRST Operator and the Superalgebra of Constraints}

Following Dirac$^{\cite{dir}}$, let us consider a Hamiltonian system with $H(p_{\alpha},q^{\alpha})$, where $\alpha=1,\ldots,M$, as Hamiltonian and $\Phi_{i}(p_{\alpha},q^{\alpha})$ as first class constraints, such that

\begin{equation}
\Phi_i (p_\alpha , q^\alpha ) = 0 , \hspace{.5in} (i= 1, \cdots,N), 
\end{equation}
and the Poisson bracket relations

\begin{equation}
 [\Phi_{i}, \Phi_{j}]_{P.B}   ={f_{ij}}^{k}\Phi_{k}~,~~~~~~
[H,\Phi_{i}]_{P.B}={\omega_{i}}^{j}\Phi_{j}~.   \label{eq:biki}
\end{equation}
Here $q^{\alpha}$ and $p_{\alpha}$ denote respectively the generalized coordinates and their canonical moments, such that

\begin{equation}
[q^{\alpha},q^{\beta}]_{P.B}=[p_{\alpha},p_{\beta}]_{P.B}=0~,~~~~~~ 
[q^\alpha,p_{\beta}]_{P.B.}={\delta^{\alpha}}_\beta~.  \label{eq:buc}
\end{equation}

For a conservative system with total energy E we may introduce

\begin{equation}
\Phi_{0}  =H - E~, 
\end{equation}
and regard $\Phi_{0}$ as an additional first class constraint. Defining

\begin{equation}
{f_{oj}}^{k}={\omega_{j}}^{k} 
\end{equation}
we can combine the two sets of Eq.(\ref{eq:biki}) into a single set with structure constants ${f_{ab}}^{c}$ antisymmetric in $a$ and $b$.

We quantize this system by turning functions of $p_{\alpha}$, $q^{\alpha}$ into hermitian operators and replacing Poisson
brackets by $(-i)$ times commutators.  The generalized coordinates satisfy

\begin{equation}
[q^\alpha,p_{\beta}]=i {\delta^{\alpha}}_{\beta},      \label{eq:balti} 
\end{equation}
and $\Phi_{a}$ become elements of the Lie algebra defined by

\begin{equation}
 [\Phi_{a},\Phi_{b}] = {f_{ab}}^{c}\Phi_{c} ,\hspace{.5in} (a=0,1,\cdots,N)~.       \label{eq:byedi}
\end{equation}
When $\Phi_{a}$ are hermitian, the structure constants are purely imaginary. If $\Phi_{a}$ are taken to be antihermitian, then ${f_{ab}}^{c}$ are real.  They also satisfy the Jacobi identity

\begin{equation}
{f_{ab}}^{c} {f_{k\ell}}^{m}+{f_{bk}}^{c} {f_{a\ell}}^{m} +{f_{ka}}^{c} {f_{b\ell}}^{m} =0~.        \label{eq:bsekiz}
\end{equation}
Note that we have not required the existence of a Killing metric. Hence indices cannot be raised or lowered and no antisymmetry of $f_{abd}$ in three lower indices is required.  In other words, the Lie algebra of constraints does not have to be semi-simple.  Moreover, the number $N$ of original constraints can be infinite as in the case of gauge theories and string theories.

In the quantized theory we require the vanishing of the matrix elements of the constraints between physical states $|\mu>$ and $|\nu>$, so that

\begin{equation}
<\mu |\Phi_{a} |\nu>= 0~. 
\end{equation}

This means that the constraints vanish weakly. The strong condition requires the vanishing of the operators themselves.

At this point we introduce Grassmann numbers $c^{a}$ as odd operators as well as their canonical conjugates $b_{a}$ such
that, in analogy to the Heisenberg relations Eq.(\ref{eq:buc}) we have the anticommutator relations

\begin{equation}
\{c^{m},c^{n}\} = \{b_{m},b_{n}\} = 0~,~~~~~~ \{c^{m},b_{n}\} =
\delta_{n}^{m}~.             \label{eq:bon}
\end{equation} 
$c^{m}$ are ghost creation operators, equal in number to the constraints $\Phi_{m}$. Out of these fermionic Heisenberg
operators which commute with the constraints we can construct another copy of the constraint Lie algebra by defining

\begin{equation}
S_{k} = -{f_{k\ell}}^{m}c^{\ell}b_{m} = c^{\ell} {f_{\ell k}}^{m}b_{m}~.            \label{eq:bonbir}
\end{equation}

Using Eqs.(\ref{eq:byedi}),(\ref{eq:bsekiz}) and (\ref{eq:bon}) we obtain

\begin{equation}
[S_{k},S_{r}] = {f_{kr}}^{m} S_{m}~.
\end{equation}

Since $b_{l}$ and $c^{m}$ can be represented by matrices, the operators $S_{k}$ form a matrix representation of the constraint Lie algebra. We also have
\begin{equation}
[\Phi_{m} , S_{n}] = 0~.
\end{equation}

It follows that the modified constraints

\begin{equation}
J_{k} = \Phi_{k}+S_{k} = \Phi_{k} - {f_{k\ell}}^{m} c^{\ell} b_{m}~,
                                          \label{eq:bondort}
\end{equation}
also obey the same Lie algebra as the $\Phi_{k}$.  The original constraints $\Phi_{k}$ and the bilinear $S_{k}$ are the analogs of the orbital angular momenta $L_{i}$ and the spin angular momenta $s_{i}$ when the constraint Lie algebra $G$ reduces to the rotation group $O(3) \sim SU(2)$. We have

\begin{equation}
[J_{k} , J_{\ell}] = {f_{k\ell}}^{m} J_{m}~,      \label{eq:bonbes}    
\end{equation}

\begin{equation}    
[J_{k},\Phi_{\ell}] = {f_{k\ell}}^{m}\Phi_{m}~,~~~~~~ [J_{k} , S_{\ell}] = {f_{k\ell}}^{m} S_{m}~.                       \label{eq:bonalti}
\end{equation}
Eq.(\ref{eq:bonbes}) tell us that the modified constraints $J_{k}$ act like total angular momentum operators, while Eq.(\ref{eq:bonalti}) express the fact that the original constraints and the "spin" operators transform like the adjoint representation of the group $G$ generated by $J_{k}$.  In the case of the $SU(2)$ group this means simply that $\mbox{\boldmath$L$}$ and $\mbox{\boldmath$S$}$ are vectors. We also find

\begin{equation}
[J_{k} , c^{\ell} ]= - {f_{kr}}^\ell c^{r} = c^{r} {f_{rk}}^{\ell}~. \label{eq:bonyedi}                   
\end{equation}

Hence the ghosts also transform like the dual of the adjoint representation of $G$.  A mirror property holds for the 
antighosts $b_{m}$ which transform like the adjoint representation of $G$, so that

\begin{equation}   
[J_{k} , b_{m}] = {f_{km}}^{\ell} b_{\ell}.   
\end{equation}

We are now ready to introduce the BRST charge operator $Q$ with the following properties

(a) - $Q$ is an odd nilpotent operator:

\begin{equation}
Q^{2} = \frac{1}{2} \{Q,Q\} = 0.
\end{equation}

(b) - $Q$ is invariant under the group $G$:

\begin{equation}
[J_{k}, Q]= 0 .
\end{equation}

(c) - $Q$ together with the modified constraints $J_{k}$ and the antighosts forms a superalgebra $S$. Hence the anticommutator of $Q$ with the antighosts must be proportional to the modified constraints. Normalizing $Q$ so that the proportionality constant is unity, we obtain

\begin{equation}
\{Q, b_{m}\} =  J_{m}~. 
\end{equation}

(d) - We shall often use a supplementary condition that $Q$ be hermitian with respect to an appropriate scalar product. This
means that we shall restrict ourselves to the unitary representations of the supergroup associated with $S$, or the
representations of the superalgebra $S$ by hermitian operators acting on the states of a definite Hilbert space.

If we assume $Q$ to be a function of $\Phi_{k}$, $c^{m}$ and $b_{n}$ we can write

\begin{equation}
Q = A c^{m} \Phi_{m} + B c^{n} S_{n}~, 
\end{equation}
since $Q$ must be odd and a scalar under $G$.  $A$ and $B$ are invariant coefficients that commute with $\Phi_{k}$, $c^{m}$ and $b_{n}$. We can take them as numerical coefficients.  Nilpotency gives

\begin{equation}
Q^{2} = A c^{m} c^{n} {f_{mn}}^{r} \Phi_r (\frac{1}{2} A - B) -
\frac{1}{2} B^{2} c^{m} c^{n} {f_{mn}}^{r} S_{r} = 0~. \label{eq:byirmiuc}  
\end{equation}

On the other hand, using the expression Eq.(\ref{eq:bonbir}) we can write

\begin{equation}
{f_{mn}}^{r}c^{m}c^{n}S_{r} = {f_{mn}}^{r} {f_{kr}}^{s} c^{m}c^{n}c^{k}b_{s}~,
\end{equation}
so that, the Jacobi identity Eq.(\ref{eq:bsekiz}) yields

\begin{equation}
{f_{mn}}^{r} c^{m}c^{n}S_{r} = 0~.
\end{equation}

Substitution of this result in Eq.(\ref{eq:byirmiuc}) shows that the condition for $Q$ to be nilpotent is given by

\begin{equation}
Q = A c^{m} (\Phi_{m} + \frac{1}{2} S_{m})~.          
\end{equation}

In order to satisfy condition (c) we calculate

\begin{eqnarray}
\{Q , b_{r}\} & =& A \{c^{m}, b_{r}\} (\Phi_{m} + \frac{1}{2} S_{m}) + \frac{1}{2}A c^{m}[S_{m} ,b_{r}] \\ \nonumber
& & = A (\Phi_{r} + S_{r}) = A  J_{r}~,
\end{eqnarray}
where we have used Eqs.(\ref{eq:bon}), (\ref{eq:bondort}) and (\ref{eq:bonyedi}). Hence the superalgebra closes if $A=1$.  We also note the relation

\begin{equation}
\{Q, c^{r}\} =  - \frac{1}{2} {f_{mn}}^r c^{m} c^{n}~.   
\end{equation}

To summarize, we have succeeded in embedding the group $G$ in a superalgebra $S$ given by

\begin{eqnarray}
\{Q , Q\} = 0~,~~~~~ \{b_m, b_n\} = 0~,~~~~~ \{Q , b_m\} =J_m~,  \nonumber 
\end{eqnarray}

\begin{eqnarray}
[J_m, Q] = 0~,~~~~~~ [J_m , b_n] = {f_{mn}}^k b_k~,  \nonumber
\end{eqnarray}

\begin{equation}
[J_m, J_n] = {f_{mn}}^k J_k~.
\end{equation}

   A realization of this superalgebra is obtained by taking $J_{m}$ to be given by Eq.(\ref{eq:bondort}) and $Q$ by

\begin{equation}
Q = c^{n}(\Phi_n+\frac{1}{2}S_n)=c^n\Phi_n +\frac{1}{2}{f_{mn}}^r c^n c^m b_r~.       \label{eq:botuzuc}
\end{equation}

Comparing with the expression Eq.(\ref{eq:bondort}) for the modified constraints $J_{m}$, note the occurrence of the factor
$\frac{1}{2}$ in front of the spin operator in Eq.(\ref{eq:botuzuc}).

Another peculiarity of the operator $Q$ is its non symmetrical appearance.  A more symmetrical form can be given if we remember that the constraints $\Phi_{n}$ are the analogs of the orbital angular momenta.  They are also functions of the canonical variables $q^{\alpha}$ and $p_{\beta}$.  Let us introduce $(N+1)$ new collective coordinates $C^{r}$ and their canonical conjugates $B_{r}$, both functions of $q^{\alpha}$ and $p_{\beta}$, that satisfy the Heisenberg relations

\begin{equation}
[B_r, B_s] = [C^{r}, C^{s}] = 0~,~~~~~~  [B_{s}, C^{r}] =\delta_{s}^{r}~,         \label{eq:botuzdort}
\end{equation}
and are chosen such that

\begin{equation}
\Phi_m = {f_{km}}^{\ell} C^{k} B_{\ell}~.
\end{equation}

Then we find that Eq.(\ref{eq:balti}) is satisfied.  In terms of the bosonic canonical pair $(B_{m},C^{n})$ and the fermionic canonical pair $(b_{m},c^{n})$ we find

\begin{equation}
J_r = {f_{mr}}^n (C^m B_{n} + c^{m}b_{n})~,           
\end{equation}

\begin{equation}
Q = {f_{mr}}^{n} c^{r} (C^{m} B_{n} + \frac{1}{2}c^{m} b_{n})~. \label{eq:botuzyedi} 
\end{equation}

These operators form a superalgebra together with $b_{r}$.  The generators of $S$ are then functions of the coordinates
$(C^{n},c^{n})$ of a superspace point and their canonical conjugate momenta $(B_{m},b_{m})$ that label a point in the
super-momentum space.

The constraint superalgebra can be enlarged through the introduction of the ghost number operator

\begin{equation}
N_{g} = c^{k} b_{k}~.
\end{equation}

Its eigenvalues give the ghost numbers.  We have

\begin{eqnarray}
[N_{g} , J_{r}] = 0 ~,~~~~~ [N_{g} , c^{r}] = c^{r}~,~~~~~ [N_{g} , b_{r}] = -b_{r}~,   \nonumber
\end{eqnarray}

\begin{equation}
[N_{g} , Q] = Q~.   
\end{equation}

Hence the constraints and modified constraints have ghost number zero.  It is $+1$ for ghosts and the BRST operator, while it is $-1$ for the antighosts $b_{r}$.

 If the superspace formulation is used we can also introduce the bosonic number operator $N_{B}$

\begin{equation}
N_B = C^{k} B_{k} 
\end{equation}
with the properties

\begin{eqnarray}
[N_B, c^r] = [N_B, N_g] = 0~,~~~~ [N_B, C^r] = C^r~,~~~~ [N_B, B_s] =-B_s~, \nonumber
\end{eqnarray}

\begin{equation}
[N_B, J_r] = 0~,~~~~~ [N_B, b_r] = 0~,~~~~~ [N_B, Q] = 0~.
\end{equation}

Finally we note that the superalgebra $(J_{r}, b_{r}, Q)$ is not semi-simple even if the group $G$ of the constraints is
semi-simple.  The hermiticity of $Q$ poses also a delicate problem.  Since $Q$ is nilpotent it can only be hermitian with respect to a scalar product with an indefinite metric.  This is often the case in relativistic theories. The infinite dimensional constraints bring additional convergence problems for the existence of $Q$.  We shall illustrate these
points by some concrete examples.

In the case of infinite algebras another difficulty may arise in the form of a central extension of the algebra. The disappearance of the central term occurs if certain restrictions are met, like the critical dimensions of strings or superstrings associated with the infinite Virasoro or super-Virasoro algebras.  In our construction the BRST charge is a nilpotent element of a super-algebra that by definition has no central extension.

\section{Properties of Physical States}

In the BRST formalism one considers BRST invariant states $|n, \alpha>$ annihilated by the operator $Q$, that are
simultaneously eigenstates of the ghost number operator $N_{g}$ with eigenvalue $n$,

\begin{equation}
Q  | n, \alpha > = 0~,            \label{eq:bkirkdort}
\end{equation}

\begin{equation}
N_g  | n, \alpha > = n  | n , \alpha >~.       \label{eq:bkirkbes}            
\end{equation}
Such states form the sector with ghost number $n$.  The property that $Q$ increases the ghost number of a state by one, coupled with its nilpotency yields a trivial solution to Eqs.(\ref{eq:bkirkdort}), (\ref{eq:bkirkbes}) of the form

\begin{equation}
 | n , \gamma >_{{\rm triv.}} = Q  | n-1 , \gamma >~.              
\end{equation}
Two solutions $|n,\alpha_{1}>$ and $|n,\alpha_{2}>$ of the same equations are equivalent if they differ by a trivial solution. Thus

\begin{equation}
| n , \alpha_{1} > =  | n, \alpha_{2} > + Q  | n -1 , \gamma >~.  
\end{equation}
Such a pair of states are regarded as belonging to the n'th cohomology class of the group $G$.

Physical states are those that belong to the zeroth cohomology class.  They are annihilated by both $Q$ and $N_{g}$.  We shall write $|\alpha>_{{\rm ph.}}$ instead of  $|0,\alpha>$.

\begin{equation}
N_g  | 0, \alpha > = \sum_n  c^{n} b_{n}  | \alpha >_{{\rm ph.}} = 0. \label{eq:bkirksekiz}
\end{equation}

Since $c^{n}$ and $b^{n}$ are canonically conjugate such a state cannot be annihilated by both ghosts and antighost operators. Hence, instead of Eq.(\ref{eq:bkirksekiz}) we can write

\begin{equation}
b_{n}  | \alpha >_{{\rm ph.}} = 0~,             \label{eq:bkirkdokuz}
\end{equation}
together with

\begin{equation}
Q  | \alpha >_{{\rm ph.}} = 0~.               \label{eq:belli}
\end{equation}
These two equations lead to

\begin{equation}
\{Q, b_n\}  | \alpha >_{{\rm ph.}} = J_{n}  | \alpha >_{{\rm ph.}} = 0~,  \label{eq:bellibir} 
\end{equation}
as well as

\begin{equation}
 c^{m} {f_{mr}}^n b_n  | \alpha >_{{\rm ph.}} = S_n  | \alpha >_{{\rm ph.}} =0~, \label{eq:belliiki}      
\end{equation}
and consequently

\begin{equation}
(J_n - S_n)  | \alpha >_{{\rm ph.}} = \Phi_n  | \alpha >_{{\rm ph.}} = 0~, \label{eq:belliuc}
\end{equation}
which is the quantum version of the original classical constraints given by the vanishing of $\Phi_{n}$.

Eqs.(\ref{eq:bkirkdokuz})-(\ref{eq:belliuc}) tell us that the physical states are invariant under the whole constraint superalgebra that we have introduced.

From a more general point of view we may consider the case of a BRST operator $Q$ that is hermitian with respect to a scalar
product defined in the vector space $|n,\alpha>$.  The metric in such a space must be indefinite, otherwise a hermitian $Q$ could not be simultaneously nilpotent.  When these two conditions are realized, taking the hermitian conjugate of Eq.(\ref{eq:bkirkdort}) with respect to the metric, we obtain

\begin{equation}
<\alpha , n  | Q = 0~,      
\end{equation}
so that $Q$ also annihilates bra states belonging to the sector with ghost number $n$.  It follows that

\begin{equation}
< \alpha , n  | (Q b_{m} + b_{m} Q)  | n, \beta > = <\alpha , n | J_{m}  | n ,\beta > = 0~. 
\end{equation}
Hence the matrix elements of the modified constraints between BRST invariant states are seen to vanish.  For $n=0$ this leads to

\begin{equation}
 _{{\rm ph.}}<\alpha  | \Phi_{m}  | \beta >_{{\rm ph.}} = 0~,            
\end{equation}
meaning that the quantum constraints vanish weakly according to Dirac's definition.
     
Because the eigenvalues of $N_{g}$ are non negative integers there is no trivial physical state.  If such a state existed, it would have been generated by $Q$ from a state $|-1,\alpha>$ with negative ghost number, but we have seen that such states are not part of our vector space.  It follows that BRST invariant states of ghost number zero are invariant under the superalgebra $S$ as well as under its Lie subgroup $G$.  For positive ghost number there will be trivial BRST invariant states and the cohomology class has to be determined at each level of ghost number $n$.

\section{Linear and Nonlinear Representations of the Constraint Superalqebra}

We start by giving a linear regular representation of the constraint superalgebra $S$.  It will be constructed out of the
adjoint representation $\Sigma_{a}$ of the constraint Lie algebra $G$. We have

\begin{equation}
{(\Sigma _a )_b}^{c} = {f_{ab}}^{c}~,\hspace{.5in} (a=0 , \ldots ,N)~, 
\end{equation}
where the coefficients $f$ are the structure constants of $G$ given by Eq.(\ref{eq:byedi}). $\Sigma_{a}$ are then $(N+1)\times (N+1)$ matrices.  We now consider the $(2N+2) \times (2N+2)$ matrices

\begin{equation}
F_{a} = \pmatrix{\Sigma_{a} & 0\cr
0  & \Sigma_{a}\cr}~,    \label{eq:bellisekiz}
\end{equation}
which form a linear representation of $G$, so that

\begin{equation}
[F_{a} , F_{b}] = {f_{ab}}^c F_c~.                     
\end{equation}

Let us introduce at this point the odd operators

\begin{equation}
\beta_a = \pmatrix{0 & \gamma\Sigma_a\cr 
               0 & 0\cr}~,                   
\end{equation}
where $\gamma^{2}=1$, $\gamma$ commutes with $\Sigma_{a}$, but anticommutes with all odd parameters  that enter in the definition of an element of the supergroup.  The inclusion of $\gamma$ is necessary for the generators $\beta_{a}$ to anticommute with odd parameters of the group.  Finally, by means of the $(N+1)\times (N+1)$ unit matrix $I$, we define

\begin{equation}
q = \pmatrix{0 & 0\cr
\gamma I & 0 \cr}~.                 
\end{equation}
These operators obey the relations

\begin{eqnarray}
\{\beta_{a} , \beta_{b}\} = 0~, \nonumber
\end{eqnarray}

\begin{eqnarray}
[F_a, \beta_b] = {f_{ab}}^c \beta_c~,  \nonumber          
\end{eqnarray}

\begin{eqnarray}
\{q, q\} = 0~, \nonumber          
\end{eqnarray}

\begin{eqnarray}
[q, F_a] = 0~,   \nonumber         
\end{eqnarray}

\begin{equation}
 \{q, \beta_{a}\} = F_a~,      
\end{equation}
showing that $F_{a}$, $\beta_{a}$ and $q$ form a linear $(2N+2)\times (2N+2)$ matrix representation of the constraint
superalgebra, with $\beta_{a}$ representing the antighosts and $q$ the BRST operator.
     
An element of the supergroup is now given by

\begin{equation}
K = {\rm exp} (F_a h^a + \beta_a \theta^a + q \varepsilon)~.   
\end{equation}

The ghost number operator $N_{g}$ is represented by the matrix

\begin{equation}
N = \pmatrix{-\frac{1}{2} I & 0\cr
                    0 & \frac{1}{2} I\cr}~.    \label{eq:baltmissekiz}
\end{equation}
Indeed we have

\begin{equation}
 [N, F_a] = 0~,~~~~ [N, \beta_a] = -\beta_a~,~~~~ [N,q] = q.         
\end{equation}

We can obtain more general matrix representations of $S$ if we represent $G$ by $k\times k$ matrices $T_{a}$.  Then, replacing $\Sigma_{a}$ by $T_{a}$ and $I$ by $I_{k}$ the $k\times k$ unit matrix in Eq.(\ref{eq:bellisekiz}) and Eq.(\ref{eq:baltmissekiz}), we immediately find a $2k\times 2k$ representation of the same superalgebra, including the ghost number.

As an example, if $G=SU(2)$ and $T_{a}= \tau_{a}$ $(a=1, 2, 3)$, the $2\times 2$ Pauli matrices, the $4\times 4$ matrix
representation of $S$ that we obtain is a $8$-dimensional ($4$ even, $4$ odd) superalgebra which is a subalgebra of the matrix algebra generated by the $4\times 4$ matrix representation of the conformal group in $(3+1)$ dimensions. 
The three operators $F_{a}$ generate $3$-dimensional rotations. The BRST operator $q$ is $\gamma$ times the time translation
(energy) operator, $\beta_{a}$ are given by $\gamma$ times the three special space-like conformal transformations (generators of constant accelerations) and finally the ghost number operator $N$ is proportional to the generator of space-time dilatation.

We now turn to a non-linear realization of $S$ in the superspace with even coordinates $C^{a}$ and odd coordinates $c^{a}$ for a point $Z$.  An infinitesimal nonlinear transformation $T_{s} Z=Z'$ of the superpoint $Z$ induces a change $\delta F$ in a function of $Z$, so that

\begin{equation}
T_s F(Z) = F (Z') = F (Z + \delta Z) = F (Z) + \delta F~, 
\end{equation}
where, in terms of the infinitesimal parameters $h^{a}$, $\theta^{a}$, $\varepsilon$ of $T_{s}$ we have

\begin{equation}
\delta F = ({\hat\Phi}_a h^a + {\hat b}_a \theta^a + {\hat q} \varepsilon) F(C^a, c^a)~.  \label{eq:byetmisbir}
\end{equation}
Here ${\hat{\phi}}_{a}$, ${\hat{b}}_{a}$, $\hat{q}$ are differential operators linear in $\partial / \partial C^{a}$ 
and $\partial / \partial c ^{a}$.

The non-linear representation will be obtained from the action of the super group on the coset of $S$ with respect to the
sub-supergroup generated by the constraints $\phi_{a}$ and the antighosts $b_{a}$ that act linearly on $Z$.  The one
odd-parameter BRST transformation that is in the coset will then have a non-linear action on $Z$.  To this end let us start from the following decomposition of a group element of $S$ in a given matrix representation

\begin{equation}
 W = T (\theta^a)~ R (h^a)~ V (\varepsilon)~, \label{eq:byetmisiki} 
\end{equation}
where

\begin{equation}
T (\theta^a) = e^{\beta_a \theta^a} = 
\pmatrix{I & \gamma \Sigma_{a} \theta^{a}\cr
               0 & I\cr}~,                 \label{eq:byetmisuc}
\end{equation}

\begin{equation}
R ( h^a) = e^{F_{a} h^{a}} = 
\pmatrix{U(h) & 0\cr 
         0 & U(h)\cr}~ ,~~~~~ (U = e^{ \sum_{a} h^{a}})~,     \label{eq:byetmisdort}       
\end{equation}

\begin{equation}
V (\varepsilon) = e^{q \varepsilon} = 
\pmatrix{I & 0\cr
        \gamma \varepsilon I & I\cr}~.    \label{eq:byetmisbes}    
\end{equation}

The group will act on a module $M$ that depends only on the parameters $C^{a}$ and $c^{a}$.

Let $m$ be a diagonal matrix that commutes with $V(\varepsilon)$. We define $M$ as the matrix obtained from $m$ through the adjoint action of the group element $\Gamma$ defined by

\begin{equation}
\Gamma = T (c^{a})~ R (\omega^{a})~ V (\zeta)~,    \label{eq:byetmisalti}      
\end{equation}
giving

\begin{equation}
M (Z) = \Gamma Z \Gamma^{-1} = T (c^a)~ R (\omega^{a}) m R^{-1}~(\omega^a) T^{-1} (c^{a})~.                
\end{equation}

The coset element $V(\zeta)$ has disappeared because it commutes with $m$. Putting

\begin{equation}
m = \pmatrix{m_{0} & 0\cr
                     0 & m_{0}\cr}~, 
\end{equation}
where $m_{0}$ does not commute with any of the $\Sigma_{a}$, we find

\begin{equation}
M (C^{a}, c^{a}) =
\pmatrix{I & \gamma \Sigma_{a}c^{a}\cr
     0 & I\cr}
\pmatrix{\Sigma_{a} C^{a} & 0\cr
                        0 & \Sigma_{a} c^{a}\cr}
\pmatrix{I & -\gamma \Sigma_{a} c^{a}\cr
             0 & I\cr}   ,    \label{eq:byetmisdokuz}                 
\end{equation}
where

\begin{equation}
\Sigma_{a} C^{a} = R (\omega^{a}) m_{o} R^{-1} (\omega^{a})~.      
\end{equation}
Thus, to each point $Z$ of the superspace $(C^{a},c^{a})$ corresponds a module $M(Z)$ given by the matrix Eq.(\ref{eq:byetmisdokuz}) that can also be written in the form

\begin{equation}
M (Z) =  
\pmatrix{\Sigma_{a} C^{a} & [\gamma \Sigma_{a} c^{a}, \Sigma_{b}
C^{b}]\cr
 0 & \Sigma_{a} C^{a}\cr}~.                      
\end{equation}

This superspace element belongs to the coset of $S$ with respect to the BRST subsupergroup $V(\zeta)$.

Under the left action of the group element $W$ given by Eq.(\ref{eq:byetmisiki}) on the group element $\Gamma$ given by 
Eq.(\ref{eq:byetmisalti}), we have

\begin{equation}
\Gamma ' = W \Gamma~.
\end{equation}

This induces the transformation

\begin{equation}
M' = M (Z') = W M (Z) W^{-1}
\end{equation}
on the module $M$.  In turn, the superspace point $Z$ is transformed into $Z'$ by the group action.  Let us consider the
action of the various sub-transformations on $Z$.  First, we can work out the action of the subgroup $G$.  We have

\begin{equation}
T_{G} M = R (h^{a}) M R (-h^{a})~,
\end{equation}
which, on using Eq.(\ref{eq:byetmisdort}) leads to

\begin{eqnarray}
M (C'^{a} , c^{a})&=&
\pmatrix{I & \gamma U \Sigma_{a}c^{a}U^{-1}\cr
     0 & I\cr}      \times  \\ \nonumber
& &\pmatrix{U \Sigma_{a}C^{a}U^{-1} & 0\cr
               0 & U \Sigma_{a}C^{a}U^{-1}\cr}
\pmatrix{I & -\gamma U \Sigma_{a}c^{a}U^{-1}\cr
         0 & I\cr} 
\end{eqnarray}
or

\begin{equation}
\Sigma_{a} C'^{a} = U \Sigma_{a} C^{a} U^{-1}~,~~~~~ \Sigma_{a} c'^{a}
= U \Sigma_{a} c^{a} U^{-1}~.      
\end{equation}

This means that both $C^{a}$ and $c^{a}$ transform under $G$ like the dual of its adjoint representation.  Hence, we have, from Eq.(\ref{eq:bonyedi})

\begin{equation}
\delta_h C^a = C^b {f_{bm}}^{a}  h^m~,~~~~~ \delta c^a = c^b {f_{bm}}^{a} h^{m}~.          
\end{equation}

Under the subsupergroup with parameters $\theta^{a}$ we have

\begin{equation}
M (C'^{a} , c'^{a}) = T (\theta^{a}) M T (-\theta^a)~.      
\end{equation}
By means of Eq.(\ref{eq:byetmisuc}), we find

\begin{equation}
\Sigma_a c'^{a} = \Sigma_a c^{a} + \Sigma_{a} \theta^a~,~~~~~~
\Sigma_{a} C'^{a} = \Sigma_{a} C^{a}~,            
\end{equation}
so that

\begin{equation}       
\delta_{\theta} C^{a} = 0~, ~~~~  \delta_{\theta} c^{a} =\theta^{a}~.                  
\end{equation}

The effect of $T(\theta)$ is to translate the odd coordinates and leave the even coordinates invariant in superspace.  Hence $G$ acts linearly on $Z$.

Turning now to the BRST transformation, we find

\begin{equation}
M' = V(\varepsilon) M V (-\varepsilon)= V T (c) V^{-1} (V F_{a}C^{a} V^{-1}) V T^{-1}(c) V^{-1}~. 
\end{equation}

Using the expression Eq.(\ref{eq:byetmisbes}) for $V$, we can write

\begin{eqnarray}
\Sigma_{b} C'^{b} = (1-\varepsilon \Sigma_{a} c^{a}) (\Sigma_{b}
C^{b}) (1-\varepsilon \Sigma_{a} c^{a})^{-1}~,  \nonumber              
\end{eqnarray}

\begin{equation}
\Sigma_b c'^{b} = (\Sigma_{b} c^{b}) (1-\varepsilon \Sigma_{a}
c^{a})^{-1}~.      
\end{equation}

  The transformation induced by the BRST operator in superspace is thus displayed to be nonlinear.  Since $\varepsilon$ is
nilpotent, we obtain

\begin{equation}
 \delta_{\varepsilon} C^{a}  = \varepsilon C^{m} {f_{mn}}^{a}c^{n}~,       \label{eq:bdoksandort}
\end{equation}

\begin{equation}
\delta_{\varepsilon} c^{a}  = -\frac{1}{2} \varepsilon {f_{mn}}^{a} c^{m} c^{n}.      
\end{equation}

The last formula is seen to be the standard transformation formula for ghosts.  The even coordinates in superspace also
transform non-linearly according to Eq.(\ref{eq:bdoksandort}).

The differential form ${\hat{q}}$ of the operator $Q$ is now obtained through the use of Eq.(\ref{eq:byetmisbir}) in the form

\begin{eqnarray}
\delta_{\varepsilon} F  &=& {\hat q}\varepsilon  F = -\varepsilon
{\hat q} F (C^{a} , c^{a}) = 
  (\delta_{\varepsilon} C^{a} \frac{\partial}{\partial C^{a}} +
\delta_{\varepsilon} c^{a} \frac{\partial}{\partial c^{a}})F  \\ \nonumber
& &= (-\varepsilon C^m {f_{mn}}^a B_a c^n -\frac{1}{2}\varepsilon {f_{mn}}^{a}c^{m}c^{n} b_{a})  
\end{eqnarray}
where we have used $B_{a}$ for $- \partial / \partial C^{a}$ and $- \partial / \partial c^{a}$to satisfy Eqs.(\ref{eq:botuzdort}) and (\ref{eq:bon}).  Finally we get

\begin{equation}
{\hat q}= c^{n} (C^{m} {f _{mn}}^{a} B_{a} + \frac{1}{2} c^{m} {f_{mn}}^{a} b_{a})~,     
\end{equation}
which is exactly the same formula as Eq.(\ref{eq:botuzyedi}) and is equivalent to the standard form given by Eq.(\ref{eq:botuzuc}).

\section{Example from Gauge Theories}

In this section we shall illustrate our general procedure by an example from a non-Abelian gauge field theory.  Let us start by reviewing the well established BRST method for an action involving gauge fixing and ghost terms.  Consider the Yang-Mills
Lagrangian based on the compact group $G$:

\begin{equation}
{\cal{L}}_{YM} = \frac{1}{4} {F_{\mu \nu}}^{a} {F_{a}}^{\mu \nu} -
\frac{1}{2} (\partial _{\mu} {A_{\nu}}^{c} -\partial_{\nu}
{A_{\mu}}^{c} + {\varphi_{ab}}^{c} {A_{\mu}}^{a} {A_{\nu}}^{b})
{F_{c}}^{\mu \nu}~, 
\end{equation}
where ${\varphi_{ab}}^c$ are the structure constants of $G$ and the coupling constant $g$ is taken to be unity.  The Lorentz metric $\eta_{\mu \nu}$ has been chosen such that

\begin{equation}
-\eta_{00} = \eta_{11} = \eta_{22} = \eta_{33} = 1~,~~~ \eta_{\mu
\nu} = 0 ~~{\rm for}~~ \mu \neq \nu.             
\end{equation}
Here ${F_{\mu \nu}}^{a}$ is an auxiliary tensor field.  The variation of the action with respect to this antisymmetrical
tensor gives

\begin{equation}
{F_{\mu \nu}}^c = {F_{\mu \nu}}^c (A) = \partial_{\mu} {A_{\nu}}^{c} - \partial_{\nu} {A_{\mu}}^{c} 
 +{\varphi_{ab}}^{c} {A_{\mu}}^a {A_{\nu}}^{b}~.                
\end{equation}

Inserting this in the Lagrangian we get the usual Lagrangian

\begin{equation}
{\cal{L}}_{YM} = -{\frac{1}{4}} {F_{\mu \nu}}^a (A)
{F_{a}}^{\mu \nu} (A)     
\end{equation}
which is invariant under the infinitesimal gauge transformation

\begin{eqnarray}
{A_{\mu}}^{'c} = {A_{\mu}}^{c} + \delta {A_{\mu}}^{c}, \nonumber        
\end{eqnarray}

\begin{equation}
\delta {A_{\mu}}^{c} = \partial_{\mu} \omega^{c} + {\varphi_{ab}}^{c}
{A_{\mu}}^{a} \omega^{b} =  D_{\mu} \omega^{c},  \label{eq:byuzdort}       
\end{equation}
where $D_{\mu}$ denotes the covariant derivative.

Variation of the action with respect to the potential yields the equation of motion

\begin{equation}
D_{\mu} {F_c}^{\mu \nu} = \partial_{\mu} {F_{c}}^{\mu \nu} +
{\varphi_{ca}}^{b} {A_{\mu}}^{a} F_b^{\mu \nu} = 0~.  \label{eq:byuzbes}
\end{equation}
When $G$ is compact, the structure constants are totally antisymmetrical in their three indices and we recover the usual
formula for the covariant derivative

\begin{equation}
D_{\mu} F_c^{\mu \nu} = \partial_{\mu} F^{\mu \nu}_{c} +
\varphi_{abc} A_{\mu}^{a} F^{\mu \nu b}~.  
\end{equation}

Under a gauge transformation the change in the field is given by

\begin{equation}
\delta {F_{\mu \nu}}^c = {\phi_{ab}}^c {F_{\mu \nu}}^{a}\omega^{b}~,         
\end{equation}
so that it transforms homogeneously.

For canonical quantization we must find the canonical conjugates to the potentials ${A_{\mu}}^{a}$.  We have

\begin{equation}
\Pi_{c}^{n}  =    \frac{\partial {\cal{L}}}{\partial(\partial_0 A_n^c)} = F_c^{n0} = E^n~,        \label{eq:byuzsekiz}
\end{equation}

\begin{equation}
\Pi_c^0  =  0~. 
\end{equation}

A covariant gauge fixing term ${\cal L}_{GF}$ is added to the Lagrangian in order to allow for covariant quantization. In terms of the auxiliary scalar fields $B_{c}$, the usual form is

\begin{equation}
{\cal{L}}_{GF} = \frac{\alpha}{2} B_c B^c - B_c \partial^\mu {A_\mu}^c.
\end{equation}

This term breaks the local gauge invariance but allows a non vanishing conjugate to ${A_{0}}^{c}$.  We have

\begin{equation}
\Pi^{0}_{c}  = \frac{\partial {\cal{L}}}{\partial(\partial_{0}A_{0}^{c})} = B_{c}~.              
\end{equation}

Thus, $B^{c}$ plays a role similar to ${F_{\mu \nu}}^{c}$. The variation with respect to it gives

\begin{equation}
\alpha B^c = \partial^{\mu} {A_{\mu}}^c~, 
\end{equation}
so that when these values of $B^{c}$ are inserted in ${\cal L}_{GF}$ we obtain

\begin{equation}
{\cal{L}}_{GF} = - {\frac{1}{2 \alpha}}  (\partial^{\mu} {A_{\mu}}^{c}) (\partial^{\nu} A_{\nu c})~, 
\end{equation}
which is the standard fixing term for Lorentz gauge.  The equation of motion Eq.(\ref{eq:byuzbes}) is now modified to

\begin{equation}
 D_{\mu} F_{c}^{\mu \nu} + \partial^{\nu} B_{c} =0~. \label{eq:byuzondort}    
\end{equation}

Since covariant quantization introduces negative probabilities and negative energies due to the presence of $A_{0}$ and
$\Pi^{0}$, these unphysical effects must be compensated by Faddeev-Popov ghost fields $c^{a}(x)$ and ${\bar{c}}_{a}(x)$
which are fermionic scalar fields and occur in the Lagrangian in the combination

\begin{equation}
{\cal{L}}_{{\rm gh.}} = -{\partial}^{\mu} {\bar{c}}_{a} D_{\mu} {c^{a}}
= -{\partial}^{\mu} {\bar{c}}_{a} ({\partial}_{\mu} c^{a} + {\varphi_{rs}}^{a} {A_{\mu}}^{r}{c^{s}})~.      
\end{equation}

The total Lagrangian is now
\begin{equation}
 {\cal{L}} ={\cal{L}}_{YM} + {\cal{L}}_{GF}+{\cal{L}}_{{\rm gh}}~.     \label{eq:byuzonalti}      
\end{equation}
Varying with respect to $\bar{c}_{a}$ we get the equation of motion

\begin{equation}
\partial^{\mu} D_{\mu} c^{a} = 0~.  \label{eq:byuzonyedi}             
\end{equation}
Varying with respect to $c^{a}$ we have

\begin{equation}
D_{\mu} \partial^{\mu} {\bar{c}}_a = \partial_{\mu} \partial^{\mu} {\bar{c}}_a + \varphi_{rsa} {A_{\mu}}^{r}
 \partial^{\mu} {\bar{c}}^{s} = 0~.    \label{eq:byuzonsekiz}         
\end{equation}
The equation of motion Eq.(\ref{eq:byuzondort}) now takes its final form

\begin{equation}
D_{\mu} F_{c}^{\mu \nu} + \partial^{\nu} B_{c} + \varphi_{abc}
(\partial^{\nu} {\bar{c}}^{a}) c^{b} = 0~.     
\end{equation}
Calling $\pi_{a}$ and ${\bar{\pi}}_{a}$ the canonical conjugates of $c^{a}$ and $\bar{c}^{a}$ respectively, we find

\begin{equation}
\pi_{a} = \partial^{0} {\bar{c}}_{a} = -\partial_{0} {\bar{c}}_{a}~,           \label{eq:byuzyirmi}                
\end{equation}

\begin{equation}
     {\bar{\pi}}^{a} = -\partial^{0} c^{a} - {\varphi_{rs}}^{a} A^{0r}c^{s} = D_{0} c^{a}~.       \label{eq:byuzyirmibir}                  
\end{equation}
It is well known that $ L$ is invariant under the BRST transformation obtained by replacing the gauge functions
$W^{a}(x)$ by $\varepsilon c^{a} (x)$ so that Eq.(\ref{eq:byuzdort}) takes the form

\begin{equation}
\delta_{\varepsilon} A_{\mu} ^{a} = \varepsilon D_{\mu}c^{a}~,        \label{eq:byuzyirmiiki}         
\end{equation}
where $\varepsilon$ is a constant Grassmann variable anticommuting with $c^{a}$ and ${\bar{c}}_{a}$. Then ${\cal L}_{YM}$ is
invariant under this transformation.  The BRST transformation leaves the auxiliary field B invariant

\begin{equation}
 \delta_{\varepsilon} B^{a} =0~,                   \label{eq:byuzyirmiuc}  
\end{equation}
and the transformation laws for the ghosts and antighosts are

\begin{equation}
\delta_{\varepsilon} {\bar{c}}_a = \varepsilon B_{a}~,       \label{eq:byuzyirmidort}
\end{equation}

\begin{equation}
\delta_{\varepsilon} c^{r} = -\frac{1}{2} \varepsilon {\varphi_{ab}}^r c^a c^b~.            \label{eq:byuzyirmibes}    
\end{equation}
Since  the term $\frac{1}{2} \alpha B^{2}$ is BRST invariant, we find

\begin{eqnarray}
\delta_{\varepsilon} {\cal{L}} & =& \delta_{\varepsilon}(-{B_{a}} \partial^{\mu} {A_{\mu}}^{a}
-{{\partial}^{\mu}}{{\bar{c}}_{a}} {D_{\mu}} {c^{a}})   \\ \nonumber
&=& -\varepsilon (B_{a} \partial^{\mu} D_{\mu} c^{a} +
\partial^{\mu} B_{a} D_{\mu} c^{a}) -\partial^{\mu}
{\bar{c}}_{a} \delta_{\varepsilon} D_{\mu} c^{a}~.          
\end{eqnarray}
Now

\begin{eqnarray}
\delta_\varepsilon D_{\mu} c^{a}  &=& -\frac{1}{2} \varepsilon \partial_{\mu} ({\varphi_{rs}}^{a} c^{r} c^{s}) + 
{\varphi_{rs}}^{a} \delta_{\varepsilon}A_{\mu}^{r} c^{s}
+{\varphi_{rs}}^{a} A_{\mu}^{r} \delta_{\varepsilon} c^{s} \\ \nonumber
& =& \varepsilon\{-\frac{1}{2} {\varphi_{rs}}^{a} ({\partial_{\mu}}
c^{r}) c^{s} - \frac{1}{2} {\varphi_{rs}}^{a} c^{r} {\partial}_{\mu}
c^{s} + {\varphi_{rs}}^{a} ({\partial_{\mu}} c^{r}) c^{s}\}  \\ \nonumber     
 & &+\varepsilon ({\varphi_{rs}}^a {\varphi_{mn}}^{r} {A_{\mu}}^{m} c^{n} c^{s}
-\frac{1}{2} {\varphi_{rs}}^{a} A_{\mu}^{r}{\varphi_{mn}}^{s} c^{m} c^{n}) \\ \nonumber
&=& \varepsilon ({\varphi_{sm}}^{a} {\varphi_{rn}}^{s} + \frac{1}{2}
{\varphi_{rs}}^{a} {\varphi_{mn}}^s) {A_{\mu}}^{r} c^{n} c^{m}  \\  \nonumber
&=& \frac{1}{2} \varepsilon ({\varphi_{sm}}^{a} {\varphi_{rn}}^{s} +
{\varphi_{sn}}^{a} {\varphi_{mr}}^{s} + {\varphi_{sr}}^{a}
{\varphi_{nm}}^{s})A_{\mu}^{r} c^{n} c^{m}~,       
\end{eqnarray}
or

\begin{equation}
\delta_{\varepsilon} D_{\mu} c^{a} = 0~, 
\end{equation}
because of the Jacobi identity.  Hence we obtain

\begin{equation}
\delta_{\varepsilon} {\cal{L}} = - \varepsilon \partial^{\mu}
(B_{a} D_{\mu} c^{a}) = - \varepsilon\partial^{\mu} \kappa_{\mu}~.       \label{eq:byuzotuzdort} 
\end{equation}
The Lagrangian changes by a total divergence, so that the action is BRST invariant.

Let us find the corresponding conserved Noether current. We have
\begin{equation}
\varepsilon j^{\mu} = \delta A_{\nu}^{a} \frac{\partial {\cal{L}}}{\partial ({\partial}_\mu {A_{\nu}}^{a})} + \delta
c^{a} \frac{\partial{\cal{L}}}{\partial (\partial_{\mu}c^{a})} +
\delta {\bar{c}}_{a} \frac{\partial{\cal{L}}}{\partial(\partial_{\mu}{\bar{c}}_a)}                
\end{equation}
where we have used Eq.(\ref{eq:byuzyirmiuc}).  The other equations Eq.(\ref{eq:byuzyirmiiki})-Eq.(\ref{eq:byuzyirmibes}) give

\begin{equation}
j^{\mu} = -(D_{\nu}c^{a})(F_{a}^{\mu \nu} + B_{a}\eta^{\mu \nu}) - \frac{1}{2} {\varphi_{m n}}^{a} 
c^{m}c^{n}\partial^{\mu}{\bar{c}}_{a} - B_{a} D^{\mu} c^{a}~,
\end{equation}

\begin{equation}
 \varepsilon \partial_{\mu} j^{\mu} = \delta_{\varepsilon}{\cal{L}} = -\varepsilon \partial_{\mu} \kappa^{\mu}~. 
                                   \label{eq:byuzotuzyedi}
\end{equation}

It follows that the conserved Noether current is given by

\begin{equation}
J^{\mu} = j^{\mu} + \kappa^{\mu} 
\end{equation}
or, using Eq.(\ref{eq:byuzotuzdort}) through Eq.(\ref{eq:byuzotuzyedi})

\begin{equation}
 J^{\mu} = -D_{\nu} c^a (F_{a}^{\mu \nu} + B_{a} \eta^{\mu \nu})
-\frac{1}{2} {\varphi_{ab}}^{k} c^{a} c^{b}\partial^{\mu}{\bar{c}}_{k}~. 
\end{equation}
We shall define the corresponding conserved charge $Q$ by

\begin{equation}
Q = \int d^{3} x~ J^{0}~,
\end{equation}
where, using Eqs.(\ref{eq:byuzsekiz}),(\ref{eq:byuzyirmi}) and Eq.(\ref{eq:byuzyirmibir}) we find

\begin{eqnarray}
J^{0}  &=& -(D_{n} c^{a}) F_{a}^{0n} - B_{a} D^{0} c^{a}
-\frac{1}{2} {\varphi_{ab}}^{k} c^{a} c^{b} \partial^{0} \bar{c}_{k} \\ \nonumber    
& =& (D_{n} c^{a}) \Pi^{n} + B_{a} {\bar{\pi}}^{a}
-\frac{1}{2} {\varphi_{ab}}^{k} c^{a} c^{b} \pi_{k}~.     
\end{eqnarray}
Doing a partial integration, we can also write

\begin{equation}
Q = \int d^{3}x~ \{ c^{a}(-D_{n} \Pi_{a}^{n} -\frac{1}{2}{\varphi_{ab}}^{k} c^{b} \pi_{k}) + {\bar{\pi}}^{a} B_{a}\}~.
                                        \label{eq:byuzkirkuc}  
\end{equation}
This is the final form of the BRST operator.  Now the classical constraints which generate gauge transformations are

\begin{equation}
L_{a} = - D_{n} \Pi^{n}_{a}~,~~~~~~   \lambda_{a} = B_{a}~.
\end{equation}

The $L_{a}$ do not commute but generate a current algebra.  The $B_{a}$ do commute.  $L_{a}$ and $B_{a}$ form an infinite Lie algebra of first class constraints.  The corresponding ghost variables are respectively $c^{a}$ and ${\bar{\pi}}_{a}$.  Since they are functions of position we must sum over the index and integrate over $\mbox{\boldmath$x$} $.  Thus the formula Eq.(\ref{eq:byuzkirkuc}) is identical with the general formula Eq.(\ref{eq:botuzuc}).  Using the canonical commutation relations it can be checked easily that $Q$ generates the transformation laws Eq.(\ref{eq:byuzyirmiiki})-
Eq.(\ref{eq:byuzyirmibes}).  On the other hand, because $Q$ has the standard form its square must also vanish. The physical states of Yang-Mills theories are therefore annihilated by $Q$.  The ghost number is given by

\begin{equation}
g = \int d_{3} x ~g^{0} = \int d_{3} x~ (\pi_{a} c^{a} +
\bar{c}_{a} \bar{\pi}^{a})~,   
\end{equation}
which is constant in time owing to the conservation of the ghost current

\begin{equation}
 g^{\mu} = (\partial^{\mu} \bar{c}_{a}) c^{a} - \bar{c}_{a} D^{\mu} c^{a}~,    
\end{equation}
which follows from the equations of motion Eq.(\ref{eq:byuzonyedi}) and Eq.(\ref{eq:byuzonsekiz}). Then, for the zero ghost number sector the modified constraints $\Phi^{a}$ coincide with the classical constraints and the usual physical states are annihilated by $Q$ and $g$.  The modified constraints $\Phi_{a}$ are given by

\begin{equation}
\Phi_{a} = -D_{n} \Pi_{a}^{n} - {\varphi_{ab}}^{k} c^{b} \pi_{k}~.
\end{equation}
More details about the BRST treatment of Yang-Mills theories with applications to their renormalizability and unitarity can be found in the excellent reviews by Baulieu$^{\cite{lba}}$ and Henneaux$^{\cite{hen}}$.  What we have just shown here is that the constraint superalgebra with even elements that represent the modified constraints $\Phi_{a}$ and $B_{a}$ as well as ghost number $g$, and with odd elements $\pi^{a}$, ${\bar{c}}_{a}$ and the BRST charge $Q$, leaves the physical states invariant. Actually, the Lagrangian ${\cal L}$ given by Eq.(\ref{eq:byuzonalti}) is invariant under a larger superalgebra that includes the anti-BRST operator $\bar{Q}$ which generates non trivial transformations for the fields $B^{a}$.  This larger invariance was discovered by Gurci and Ferrari$^{\cite{gfe}}$, and Ojima$\cite{oji}$ and further developed and formalized by Baulieu and Thierry-Mieg$^{\cite{btm}}$ and Alvarez-Gaum\'{e} and Baulieu$^{\cite{agb}}$.

\section{Example from String Theories}

As a last example we shall display the infinite constraint superalgebra for a bosonic string.  Consider the loop algebra
generated by the differential operators

\begin{equation}
L_{m} = - z^{m+1} \frac{d}{dz}~ , 
\end{equation}
which obey the infinite Lie algebra

\begin{equation}
[L_{m} , L_{n}] = (m-n) L_{m+n} = (m-n) \delta^{r}_{m+n} L_{r}~. 
\label{eq:byuzkirkdokuz}
\end{equation}
This is a Virasoro algebra without central extension, also called a loop algebra.  Its structure constants are given by

\begin{equation}
{f_{mn}}^{r} = (m-n) {\delta^{r}}_{m+n} = -{f_{nm}}^{r}~. 
\end{equation}

Introduce conjugate Grassmann numbers $c^{m}$ and $b_{n}$ such that

\begin{equation}
\{c^{m} , c^{n}\} = \{ b_{m} , b_{n}\} = 0~,~~~~~~ \{c^{m} , b_{n}\} = 
\delta_{n}^{m}~. 
\end{equation}

A finite  subalgebra consists of the dilation operator

\begin{equation}
L_{0} = -z \frac{d}{dz}~, 
\end{equation}
the translation operator

\begin{equation}
L_{-1}= -\frac{d}{dz}~, 
\end{equation}
and the translation operator for the inverse $z^{-1}$, namely

\begin{equation}
L_{1} = -z^{2} \frac{d}{dz}~.
\end{equation}
$L_{0}$, $L_{\pm 1}$ generate the group $SL(2,R)$ and the Virasoro algebra may be regarded as its affine extension.
 
A more general form of the Virasoro algebra without central extension can be constructed by means of the harmonic oscillator 
operators $a$ (multiplication by $z$) and $a^{\dagger}$ (represented by the differential operator $-d / dz$)

\begin{equation} 
L_{m} = a^{m+1} a^{\dagger} + h(m+1) a^{m}~,  \label{eq:byuzellibes}
\end{equation}
which also obey Eq.(\ref{eq:byuzkirkdokuz}).  Then $L_{0}$ and $L_{\pm 1}$ correspond to the well known Dyson representation$^{\cite{dys}}$ of $SL(2,R)$ with Casimir invariant

\begin{equation}
C = L_{0}^{2} - \frac{1}{2} \{L_{1} , L_{-1}\} 
\end{equation}
equal to $h(h+1)$.

Now, construct the "spin" operators

\begin{equation}
S_{n} = c^{m} {f_{mn}}^{r} b_{r} = (m-n)c^{m} b_{m+n}~. 
\end{equation}
  
The modified constraints are
\begin{equation}
\Phi_{n} = L_{n} + S_{n} = L_{n} + (m-n)c^{m} b_{m+n}~,   \label{eq:byuzellisekiz}
\end{equation}
while the BRST operator is given by

\begin{equation}
Q = c^{m} (L_{n} + \frac{1}{2} S_{n}) = c^{n}L_{n} + 
\frac{1}{2} (m-n)c^{n}c^{m}b_{m+n}~.
\end{equation}

The associated superalgebra is again given by $\Phi_{n}$, $b_{n}$ and $Q$.  The ghost number operator is

\begin{equation}
g = c^{n} b_{n}~. 
\end{equation}
               
Note that for the representation Eq.(\ref{eq:byuzellibes}) $L_{-n}$ is different than $L_{n}^{\dagger}$. If we have a representation for which

\begin{equation}
L_{-n} = L_{n}^{\dagger}~,~~~~~~   L_{0} = L_{0}^{\dagger}~, 
\end{equation}
then the Virasoro algebra Eq.(\ref{eq:byuzkirkdokuz}) defined for $L_{0}$ and $L_{n}$ ($n>0$) defines the whole algebra.  In that case it is enough to impose constraints for $n \ge 0$ only. When ghost variables $c^{n}$ and $b_{m}$ are introduced, we must also require

\begin{equation}
c^{-m} = (c^{m})^{\dagger}~,~~~~~ b_{-m} = b_{m}^{\dagger}~. 
\end{equation}           
These conditions lead to the hermiticity of $Q$.

In string theory the Virasoro operators are the moments of the two-dimensional energy-momentum tensor.  They can be 
expressed in terms of oscillator mode operators $\alpha_{n}$ in the form$^{\cite{gsw}}$

\begin{equation}
L_{m} = \frac{1}{2} \sum_{n=-\infty}^{\infty} \alpha_{m-n}^{\mu} \alpha_{n \mu}~,~~~ (\mu = 0 , 1, \ldots, D-1)~, 
\end{equation}
where

\begin{equation}
\alpha_{-m}^{\mu} = \alpha_{m}^{\mu \dagger}~,~~~~~  [\alpha_m^{\mu} , \alpha_{n}^{\nu}] = m \delta^{0}_{m+n} 
\eta^{\mu \nu}~.
\end{equation}
               
Physical states are annihilated by $L_{m}$ ($m>0$) and $L_{0-\gamma}$ with $\gamma$ being a constant and $L_{0}$ normal
ordered:

\begin{equation}
L_{0} = \frac{1}{2} \alpha_{0}^{2} + \sum_{n=1}^{\infty} \alpha_{-n}^{\mu} \alpha_{n \mu}~. 
\end{equation}
     
The hermiticity conditions are now satisfied, but there is a price: for this string representation the Virasoro algebra 
develops a well known anomaly term and takes the form

\begin{equation}
[L_{m} , L_{n}] = (m-n) L_{m+n} + \frac{D}{12} m (m^{2}-1)
\delta^{0}_{m+n}~. 
\end{equation}

This modified algebra has the same $SL(2,R)$ subalgebra generated by $L_{0}$ and $L_{\pm 1}$.

The spin algebra develops a different anomaly term, namely

\begin{equation}
[S_{m} , S_{n}] = (m-n) S_{m+n} + \frac{1}{6} m(1-13m^{2})
\delta^{0}_{m+n}~. 
\end{equation}
         
The modified constraints $\Phi$ in Eq.(\ref{eq:byuzellisekiz}) will then satisfy the anomaly free algebra

\begin{equation}
[\Phi_{m}, \Phi_{n}] = (m - n) \Phi_{m + n}~, 
\end{equation}
provided that the total anomaly vanishes.  This anomaly must include an additional term $2\lambda m \delta_{m+n}$ induced 
in $L_{0}$ by normal ordering.  Hence we must have

\begin{equation}
\frac{D}{12} m (m^{2} - 1)+ \frac{1}{6} m (1 - 13 m^{2}) + 2
\lambda m = 0~,
\end{equation}
which gives

\begin{equation}
D = 26 ~,~~~~~~~ \lambda = 1.
\end{equation} 
           
Thus in the string representation of the Virasoro algebra the modified constraints form a Lie algebra embedded in the 
superalgebra generated by $\Phi_{m}$, $b_{m}$ and $Q$, only if the string is in 26 dimensions with intercept $\lambda$ 
equal to one.  The physical states are invariant under this constraint superalgebra which fails to exist in other 
dimensions for the bosonic string.

In mathematical terms, the cohomology for the Virasoro algebra has been rigorously solved by Frenkel, Garland 
and Zuckerman$^{\cite{zuc}}$.  Hence, the ghost-free and manifestly unitary treatment of the bosonic string on
the light cone is now fully justified.

\section{Further Outlook}

We have established the existence of a super algebra in which the Lie algebra of the first class constraints is 
embedded. The odd generator that commutes with the constraints is the BRST operator.  Its form follows from a non linear coset representation of the superalgebra. The superalgebra exists for all Yang-Mills theories and for $26$-dimensional bosonic strings.

In Yang-Mills theories two BRST operators $Q$ and $\bar{Q}$ can be introduced.  This suggests that the constraints can 
be embedded in larger superalgebras.  Such formal possibilities have already been explored$^{\cite{bow}}$. It remains to show their utility in the covariant quantization of gauge or string theories, especially when the superalgebra is a simple one.

When constraints are second class, Poisson brackets are replaced by Dirac brackets.  Can such constraints be also embedded in a superalgebra?  The resolution of the problem may have bearing on the covariant quantization of supergravity and super particle theories.

In the more general case of both bosonic and fermionic constraints that form a superalgebra, the latter can be regarded as a 
subalgebra of a larger superalgebra involving both bosonic and fermionic ghosts and BRST operators. This possibility is 
realized for the superstring. The embedding super algebra then exists only is the critical dimension $D=10$. A complete 
classification of all conditions under which the constraint superalgebra exists remains to be worked out.

\end{document}